\documentclass[floatfix,aps,showpacs,preprint,nofootinbib,preprintnumbers]{revtex4}
\usepackage{graphicx}
\begin{document}

\title{Clustering of Primordial Black Holes. II.  Evolution of Bound Systems}
\author{James R. Chisholm}
\affiliation{Southern Utah University, Cedar City, UT 84720; chisholm@suu.edu}

\date{\today}

\begin{abstract}
Primordial Black Holes (PBHs) that form from the collapse of density perturbations are more clustered than the 
underlying density field.  In a previous paper, we showed the constraints that this has on the prospects of 
PBH dark matter.  In this paper we examine another consequence of this clustering: the formation of bound systems
of PBHs in the early universe.  These would hypothetically be the earliest gravitationally collapsed structures, forming
when the universe is still radiation dominated.  Depending upon the size and occupation of the clusters, PBH merging
occurs before they would have otherwise evaporated due to Hawking evaporation.

\end{abstract}


\maketitle

\section{Introduction}

Primordial black holes (PBHs) are a unique probe of cosmology, general relativity, 
and quantum gravity.  PBHs form from the gravitational collapse of density perturbations 
that are of order unity on the scale of the cosmological horizon \cite{zeldovich,hawking1}.  
Measurements of the cosmic microwave background (CMB) anisotropy \cite{WMAP1} imply that density
perturbations at the time of decoupling are much smaller ($\delta_H \approx 10^{-5}$).  As such, 
PBH formation will be cosmologically negligible during and beyond this era.  Less constrained are
the conditions in the early universe before decoupling, and we cannot preclude the existence of much
larger density contrasts which could have formed PBHs.

PBHs would be the first gravitationally collapsed objects in the universe.  As clustering
is ubiquitous in other, observed gravitationally collapsed systems (galaxies, clusters of 
galaxies, superclusters, etc), it will be no different for PBHs.  In a previous paper \cite{chisholm}, 
we derived the basic properties of PBH clustering and what consequences it had on the viability of 
PBHs as dark matter (DM).  As PBHs are created with an isocurvature component, this constrained the 
mass and abundance of PBHs if they are to serve as DM.

The aim of this work is to continue the analysis of PBH clustering, particularly to investigate the
creation and behavior of bound systems, or clusters, of PBHs.  In Sections~\ref{basics} and \ref{clustering} 
we describe general properties of PBHs and their clustering, respectively, that will be used throughout the paper.  
The main results of the paper are presented in Section~\ref{evolution} and conclude in Section~\ref{conc}.  We use
units throughout such that $c=1$.

%%%%%%%%%%%%%%%%%%%%%%%%%%%%%%%%%%%%%%%%%%%%%%%%%%%%%%%%%%%%%%%%%%%%%%%%%%%%%%%%%%%%%%%%%%%%%%%%%
%%%%%%%%%%%%%%%%%%%%%%%%%%%%%%%%%%%%%%%%%%%%%%%%%%%%%%%%%%%%%%%%%%%%%%%%%%%%%%%%%%%%%%%%%%%%%%%%%
\section{PBH Basics}\label{basics}
%%%%%%%%%%%%%%%%%%%%%%%%%%%%%%%%%%%%%%%%%%%%%%%%%%%%%%%%%%%%%%%%%%%%%%%%%%%%%%%%%%%%%%%%%%%%%%%%%
%%%%%%%%%%%%%%%%%%%%%%%%%%%%%%%%%%%%%%%%%%%%%%%%%%%%%%%%%%%%%%%%%%%%%%%%%%%%%%%%%%%%%%%%%%%%%%%%%

A black hole of mass $M$ has a Schwarzschild radius $R_S = 2GM = \frac{2M}{M_P^2}$.  Throughout
we assume that any PBHs have negligible angular momentum and electric charge, unless otherwise
noted.  It is assumed that PBHs are formed at a fraction $f$ of the horizon mass
\begin{equation}\label{pbh_horizon_mass_relation}
M_{PBH} = f M_H, 
\end{equation}
\begin{equation}\label{horizonmass}
M_H(t) = M_P \left(\frac{t}{t_P}\right) = (2\times10^{5} M_\odot) \left(\frac{t}{1 \textrm{s}}\right)
\end{equation}
in the radiation dominated era.  An important parameter in the clustering and abundance of PBHs is 
\begin{equation}\label{nu}
\nu = \delta_c/\sigma_{rad}(r_H)
\end{equation}
which measures the height of the density peak at the threshold for PBH formation in units of the variance
of the radiation density perturbation $\sigma_{rad}$ smoothed over the horizon radius $r_H$ when that perturbation
enters the horizon.  We will examine the consequences of clustering for a range of values of $\nu$, $f$ and $M_H$, 
as there is considerable variation in their respective values.  The horizon fraction $f$ depends upon details of the 
radiation perturbation profile and equation of state (see a discussion in \cite{chisholm}).  The horizon mass $M_H$ and $\nu$ 
are determined primarily by the power spectrum peak location and value, respectively, the details of which will depend
upon the exact PBH formation model invoked.  Possibilities include: models where PBHs are produced at a particular scale in 
pre/reheating during/after multi-field (hybrid) inflation \cite{GBLW, yokoyama,malik,yamaguchi,STBK};
models with a peak at a specific mass/length scale due to a drop in pressure during a phase transtion (see \cite{jedamzik, ichiki} for examples relating
to the QCD phase transition); generic inflationary models with blue spectra \cite{kim1, gilbert}. 
In all of these models, the PBH mass function is strongly peaked, and so for simplicity we take a monochromatic mass function.

The fluctuation of a density field (either for radiation or PBHs) is defined in terms of the mean density,
\begin{equation}\label{delta}
 \delta = \frac{\rho-\bar \rho}{\bar \rho}.
\end{equation}

PBH production is exponentially suppressed ($e^{-\frac{\nu^2}{2}}$); exact expressions for the initial density depend 
upon $\nu$ and assumptions as to how the overdense region is defined.  Assuming the peaks are Gaussian, the probability
of a given point in the smoothed (over the horizon scale) density field falling above the PBH formation threshold is
\begin{equation}
\beta = \int_{\delta_c}^{1} \left(2 \pi \sigma_{rad}^2(r_H)\right)^{-1/2} \exp \left(- \frac{\delta^2}{2 \sigma_{rad}^2(r_H)}\right) d\delta,
\end{equation}
which will roughly be equal to the PBH number density at formation across many horizon-sized regions.  In the limit that $\nu \gg 1$, we obtain an approximate expression by taking the upper limit to infinity,
\begin{equation}
\label{beta}
\beta = \textrm{erfc} \left(\frac{\nu}{\sqrt{2}}\right) \approx \left(\sqrt{\frac{2}{\pi}} \nu^{-1}\right) e^{-\nu^2/2}.
\end{equation}
More precise calcuations (assuming, e.g., that the given point is not just above threshold but also is a maximum) for a power law (of slope $n$) radiation spectrum \cite{BBKS, GLMS} give different prefactors for the above
exponential.  We define the function $N_*(\nu)$ by
\begin{equation}
\beta = N_*(\nu) e^{-\nu^2/2},
\end{equation}
with values for different models being 
\begin{eqnarray}
\label{nstar}
&& N_*(\nu) = \left\{\begin{array}{cl}
\sqrt{\frac{2}{\pi}}\nu^{-1} & \textrm{erfc approximation}\\
\frac{1}{\sqrt{2\pi}} \left(\frac{n+3}{6}\right)^{3/2} \left(\nu^2-1\right) & \textrm{BBKS}\\
\end{array} \right\}.
\end{eqnarray}
The initial PBH density is then 
\begin{equation}
n_{PBH} = \frac{\beta}{V_H} = \frac{N_*(\nu) e^{-\nu^2/2}}{V_H}
\end{equation}
in terms of the Hubble volume at formation
\begin{equation}
 V_H=(2\pi)^{3/2} R_H^3.
\end{equation}

Due to quantum effects \cite{hawking2}, a BH of mass $M$ will emit particles as a blackbody 
with temperature $T_{h}$ given by
\begin{equation}
T_{h}(M) = \frac{1}{8\pi G M} = \frac{M_P^2}{8 \pi M} \approx 10^{22} \left(\frac{M}{1 \textrm{g}}\right)^{-1} \textrm{eV}.
\end{equation}
As the temperature is inversely proportional to the mass, this is unobservable for a one 
solar mass BH ($T_h(M_\odot) \approx 62$ nK), but cannot be neglected in the mass range
of PBHs.  This emission also corresponds to a mass loss for the PBH, 
\begin{equation}
\dot M = - L_h = - \sigma_{SB}^* T_h^4 (4 \pi R_s^2) = -\frac{\alpha(M)}{M^2},
\end{equation}
where $\sigma_{SB}^*$ is the effective Stefan-Boltzmann constant and is proportional to the effective number
of relativistic degrees of freedom in the emitted particles.  PBHs therefore have a finite
liftime, after which they would have emitted their entire rest mass, given by
\begin{equation}\label{lifetime}
\tau = \frac{M_0^3}{3\alpha(M_0)} = (10^{-26} \textrm{s}) \left(\frac{M}{1 \textrm{g}}\right)^3
\end{equation}
As the lifetime scales with $M^3$, there is a threshold mass above which holes will not have evaporated by 
the present day ($t_0$).  This threshold mass $M_*$ is given by
\begin{equation}
M_* \approx (4\times10^{14} \textrm{g}) \left[\left(\frac{\alpha(M_*)}{6.94\times10^{25}\textrm{g}^3/\textrm{s}}\right)\left(\frac{t_0}{4.4\times10^7 \textrm{s}}\right)\right]^{1/3}.
\end{equation}
Given the uncertainties in $\alpha$ and $t_0$, a threshold mass of $M_* \sim 10^{15}$ g is typically quoted in the literature.

A large enough abundance of PBHs with $M \approx M_*$ will produce a number of observable effects
through their evaporation in the current day.  They would contribute to cosmic rays \cite{macgibbon3}, the
$\gamma$-ray background \cite{page3,halzen}, 511 keV emission due to positron annihilation in the galactic center 
\cite{okele} or be the cause of short duration gamma ray bursts \cite{green3, cline}.
Observations (or the lack thereof) of PBHs evaporating today depend critically upon not only the 
number density of PBHs present today $n_{PBH}(t_0)$, but also upon how clustered they are within
within the galaxy.  Assuming an isothermal halo model, the effective number density is $\zeta n_{PBH}(t_0)$
where $\zeta$ is the local density enhancement factor \cite{macgibbon3, halzen, page3} and ranges from
$10^5 - 10^7$.

PBHs with $M < M_*$ would have evaporated by the present day.  The main mechanism for ``observing'' PBHs in cosmology is through their
Hawking radiation.  In the absence of a direct detection, the main utility of PBHs is to 
set limits of PBH abundance at various times given a non-detection.  Though, PBHs have also been
invoked to explain baryogenesis \cite{barrow2}, reionization \cite{gibilisco} 
and as a solution to the magnetic monopole problem \cite{stojkovic}.

Evaporating PBHs have their most dramatic effect during the era
of BBN, where Hawking radiation can alter the entropy per baryon and light element abundances
\cite{vainer1, vainer2, lindley1}.  Therefore, the success of BBN implies an upper limit to the
number of PBHs evaporating at that time.  

Combining Equations
(\ref{horizonmass}), (\ref{pbh_horizon_mass_relation}) and (\ref{lifetime}) gives the
relation
\begin{equation}
\tau(t) = \frac{f^3 M_P^3}{3 \alpha} \left(\frac{t}{t_P}\right)^3,
\end{equation}
the lifetime $\tau$ of a PBH created at a time $t$.  What this allows one to do is use 
information from a ``late epoch'' (time $\tau$) to examine conditions at an ``early epoch'' 
(time $t \ll \tau$).  In the above example, $\tau \sim t_{BBN}$, and the limits on initial
PBH abundance from BBN imply $\beta < 10^{-16}$ for $M_{PBH}$ between $10^{9}$ g and $10^{15}$ g.

This relation depends critically upon the PBH only losing mass through evaporation, and not
gaining mass in any way (accretion or merging).  Should this not be the case, the lifetime $\tau$
is no longer given by the initial PBH mass, and the link between late epoch and early epoch
is broken.  Instead, the energy in PBHs that would have evaporated away can now linger for
longer periods of time.  Since $\tau \propto M^3$, the merging of two equal mass BHs will
result in a BH with a lifetime 8 times as long.  If this merging can continue, then there is a 
greater chance of PBHs produced in the early universe still existing today.

%%%%%%%%%%%%%%%%%%%%%%%%%%%%%%%%%%%%%%%%%%%%%%%%%%%%%%%%%%%%%%%%%%%%%%%%%%%%%%%%%%%%%%%%%%%%%%%%%
%%%%%%%%%%%%%%%%%%%%%%%%%%%%%%%%%%%%%%%%%%%%%%%%%%%%%%%%%%%%%%%%%%%%%%%%%%%%%%%%%%%%%%%%%%%%%%%%%
\section{PBH Clustering}\label{clustering}
%%%%%%%%%%%%%%%%%%%%%%%%%%%%%%%%%%%%%%%%%%%%%%%%%%%%%%%%%%%%%%%%%%%%%%%%%%%%%%%%%%%%%%%%%%%%%%%%%
%%%%%%%%%%%%%%%%%%%%%%%%%%%%%%%%%%%%%%%%%%%%%%%%%%%%%%%%%%%%%%%%%%%%%%%%%%%%%%%%%%%%%%%%%%%%%%%%%

Here we briefly summarize the results of \cite{chisholm}.  Defining the radiation and PBH two point correlation 
functions $\xi(r)$ and power spectra $P(k)$ under the assumption of spherical symmetry:
\begin{equation}
\xi_{rad}(r) = \langle \delta_{rad}(x) \delta_{rad}(x+r) \rangle 
\end{equation}
\begin{equation}\label{xipbh}
\xi_{PBH}(r) = \langle \delta_{PBH}(x) \delta_{PBH}(x+r) \rangle 
\end{equation}
Using a Gaussian window function $W_k(k r_H)$, we will work with smoothed versions of the correlation functions over 
the horizon size $r_H$:
\begin{equation}
\xi_{rad}(r) = \frac{V}{2\pi^2} \int dk k^2 P_{rad}(k)\frac{\sin (kr)}{kr} |W_k(k r_H)|^2
\end{equation}
\begin{equation}
\xi_{PBH}(r) = \frac{V}{2\pi^2} \int dk k^2 P_{PBH}(k)\frac{\sin (kr)}{kr} |W_k(k r_H)|^2.
\end{equation}
It is useful to redefine the radiation field correlation function in a normalized fashion
\begin{equation}
 w(r)=\frac{\xi_{rad}(r)}{\xi_{rad}(0)}=\frac{\xi_{rad}(r)}{\sigma^2_{rad}(r_H)}
\end{equation}
so that $w(0)=1$.  

As PBHs form at the peaks of the radiation density field (where a radiation perturbation $\delta > \delta_C$), the PBH correlation
function is itself a function of the radiation correlation function.  In general, for Gaussian perturbations, the peaks of a density field
are more clustered than the density field itself; this was first illustrated by Kaiser \cite{kaiser} comparing rich clusters of galaxies
(Abell clusters) to just galaxies.  This will be the case for PBHs too; PBHs will be more clustered than the underlying radiation field ($\xi_{PBH} >> \xi_{rad}$).

As originally shown in \cite{PolWis, jensen}, the two-point correlation function in the limit of large $\nu$ becomes \cite{chisholm}
\begin{equation}
\label{politzer}
1 + \xi_{PBH}(r) = \exp(\nu^2 w(r)).
\end{equation}
Note that the radiation density field must possess large (order unity) fluctations at the horizon scale
at which PBHs are formed; this implies that the PBH ``fluctuations'' must be even larger (exponentially so).  Their 
evolution is thus intrinsically nonlinear.  As the universe expands and nearby PBHs enter the same Hubble volume, they
immediately detach from the general expansion and collapse to form bound clusters.  Their dynamics will be determined by 
the size of the cluster and typical separation within the cluster.  As PBHs form in 
(initially) separate horizon volumes, the horizon distance si the smallest possible distance scale over which PBHs could be
correlated.  As we will show later, the closest pairs of PBHs will form withing a few horizon distances from each other.

The exact details of the PBH clustering depend upon the underlying density field, represented here by the exact form
of $w(r)$.  Given $\xi_{PBH}(r)$ we can transform to find the PBH power spectrum $P_{PBH}(k)$.  Because PBHs are 
discrete objects, the limit of the power spectrum as $k \rightarrow 0$ is a constant and related to the number density
of PBHs.  The power spectrum for a group of $N$ uniformly randomly distributed objects over a volume $V \gg V_H$ is 
$1/N=(n_{PBH}V)^{-1}=\beta^{-1}$.  Our $P_{PBH}(0) \neq \beta^{-1}$, indicating that the PBHs are distributed as clusters
of objects with mean occupation number
$N_c=P_{PBH}(0)\beta$.  Due to the exponential enhancement of the correlation function, $N_c$ scales exponentially with $\nu$ as well.  The
exact relation will depend upon the assumed form of $w(r)$, and thus the assumed power spectrum $P_{rad}(k)$.  In 
\cite{chisholm}, we assumed a Gaussian spike and obtained $N_c \sim N_*(\nu)e^{\nu^2/4}$.

%%%%%%%%%%%%%%%%%%%%%%%%%%%%%%%%%%%%%%%%%%%%%%%%%%%%%%%%%%%%%%%%%%%%%%%%%%%%%%%%%%%%%%%%%%%%%%%%%
%%%%%%%%%%%%%%%%%%%%%%%%%%%%%%%%%%%%%%%%%%%%%%%%%%%%%%%%%%%%%%%%%%%%%%%%%%%%%%%%%%%%%%%%%%%%%%%%%
\section{PBH Cluster Evolution}\label{evolution}
%%%%%%%%%%%%%%%%%%%%%%%%%%%%%%%%%%%%%%%%%%%%%%%%%%%%%%%%%%%%%%%%%%%%%%%%%%%%%%%%%%%%%%%%%%%%%%%%%
%%%%%%%%%%%%%%%%%%%%%%%%%%%%%%%%%%%%%%%%%%%%%%%%%%%%%%%%%%%%%%%%%%%%%%%%%%%%%%%%%%%%%%%%%%%%%%%%%
\subsection{Literature}

PBH clusters were first explored in \citet{freese}, where PBHs in the mass range 
$ M \in \left[10^{15} \textrm{g}, 10^{33} \textrm{g}\right]$ form clusters after matter-radiation equality.  The collapse of 
baryons onto these clusters creates explosions that act as the
seeds of large scale structure in the Ostriker-Cowie \cite{OstrikerCowie} model.

As shown in the previous section, PBH perturbations enter the horizon with a very large
amplitude ($\sigma_{PBH} \sim e^{\nu^2/2}$).  It is therefore no longer valid to treat
their evolution using linear perturbation theory, as one is able to do for other forms
of CDM.  Instead, we examine the sub-horizon evolution of the PBH population as an N-body
problem.  As noted earlier, using numerical simulations to examine PBH formation is difficult
given the small numbers of them that form.  This is more so true should one want to examine
their subsequent non-linear evolution after creation.

We instead appeal to previous work done in the context of DM halo formation and N-body
simulations to examine the PBH population behavior.  Being non-relativistic, PBHs will 
cluster hierarchically (just as CDM); creating bound systems that get incorporated
into larger ones.  The internal dynamics of these systems are determined solely by gravitational
clustering, analagous to other gravitationally bound systems such as star clusters and galaxies.
For this, we are aided by the work done in the context of
studying more massive black holes in globular clusters \citep{sigurdsson} and galaxies \citep{begelman}.
In those cases, gravitational interactions tend to either produce bound pairs or ejections,
rather than BH coalescence \citep{saslaw}.  What occurs in the case of PBHs depends upon
how many ($N$) form in a ``PBH cluster'' and what their initial separations $D$ are.

We shall discuss the clustering in a hierarchical manner as well, first discussing the small
scale dynamics (PBH binary formation and evolution), then the larger scale (cluster relaxation,
ejection, and evaporation).

\subsection{PBH Binaries}
PBHs will form at rest with respect to the background expansion.  A PBH perturbation entering
the horizon implies that PBHs are able move appreciably in response to the gravity of other PBHs. 
The cluster as a whole begins with no kinetic energy and, properly defined, negative potential energy,
for negative total energy.  Starting at rest also implies no net angular momentum (either orbital or 
in BH spin) for the cluster, and further no net angular momentum for any subset of the cluster.
The first physical process to consider is direct PBH merging due to a head-on collision between two
PBHs in a cluster (to create a PBH with sum of the two initial PBHs' masses).  For the simplest case
of $N=2$, this type of merging is automatic, as there are no significant tidal forces to keep the PBHs from 
colliding\footnote{Insignificant tidal forces would arise from density perturbations in the background radiation field.  While a ``nearby'' perturbation might be just
below threshold for PBH creation, it will decay upon horizon entry.}.  For any larger cluster ($N > 2$), the situation becomes more complicated.  In this case, any pair (or subset)
of PBHs will have non-vanishing orbital angular momentum with respect to any other member of the cluster,
and tidal forces suppress the head-on merging of PBHs.  Rather, close encounters between PBHs are more 
prone instead to create PBH binary systems through a dissipative process via emission of gravitational waves.  
We will address more carefully the conditions for binary formation in Sec.\ref{PBHCO}.

Thus, we expect binary PBH formation to be common within the cluster, and we examine the evolution
of PBH binaries.  That PBHs would form binaries in the RD era was for studied in the context of 
MACHO PBHs of around a solar mass \citep{nakamura, ioka1, hiscock, ioka2} due to the expected gravitational
wave emission.  There are a number of other physical processes other than gravity that may affect
the PBH binaries.  To estimate the magnitude of each effect, we first assume that two PBHs
of identical mass $M$ are in a circular orbit of radius $D$ (a separation of $2D$).
In the initial limit that the PBHs are well separated compared to their
Schwarzschild radii ($R_S = 2GM \ll D$) and also moving non-relativistically ($v<<1$), 
the Newtonian force is

\begin{equation}
F_{grav} = \frac{GM^2}{4 D^2} = \frac{M_p^2}{16} \left(\frac{R_S}{D}\right)^2
\end{equation}

The orbital timescale (``period'') is therefore
\begin{equation}
t_{orbit} = 4 \pi D \left(\frac{D}{GM}\right)^{1/2} = 4\sqrt{2} \pi D \left(\frac{R_S}{D}\right)^{-1/2} 
\end{equation}

We next examine three mechanisms that afect the orbital motion of a PBH binary: background radiation drag,
Hawking radiation pressure, and gravitational wave emission.

\subsection{Background Radiation Drag} PBHs moving through the background will 
accrete radiation, and this will have a compensating effective drag force on the PBHs.  
Considering the radiation to be a perfect fluid with sound speed $c_s = c/\sqrt{3}$, 
the accretion rate for an individual PBH will be:

\begin{equation}
\dot{M}_{bondi} = \rho_r \pi R_{acc} \sqrt{v^2 + c_s^2},
\end{equation}
where $v$ is the PBH physical (not comoving) velocity with respect to the rest frame of
the radiation and $R_{acc}$ is the accretion radius, defined as
\begin{equation}
R_{acc} = \frac{2 G M}{v^2 + c_s^2}
\end{equation}
Due to the radiation sound-speed being so high, the accretion radius can be approximated by its non-relativistic
limit of $R_{acc} = 3 R_S$.  

Thus, the drag force $F_{drag} \sim \dot{M}_{bondi} v \sim M a$, or
\begin{equation}
F_{drag} = \frac{36 \pi}{\sqrt{3}}M^2 \left(\frac{\rho_r}{M_p^4}\right)
\end{equation}
giving the time-scale for drag
\begin{equation}
t_{drag} = \frac{v}{a} = \frac{M}{\dot{M}_{Bondi}} = \frac{1}{3^{3/4}\cdot4\pi} \left(\frac{\rho_r}{M_P^4}\right)^{-1} M^{-1} = \frac{\sqrt{3}}{18} \left(\frac{M_p^2}{\rho_r}\right)\frac{1}{R_S}
\end{equation}
Note that this is also the mass-doubling time for the PBH due to accretion.  In the
radiation dominated era we can rewrite this as
\begin{equation}
t_{drag} = \frac{\sqrt{3}}{18} \left(\frac{8\pi}{3}(4t)^2\right)\frac{1}{R_S} = \frac{16}{\sqrt{3}} \left(\frac{t}{R_S}\right)t.
\end{equation}
Since $R_S < t_H = 2t$ at creation, $t_{drag} > t$ and therefore radiation drag is negligible \citep{carr8}.

\subsection{Hawking Radiation Pressure} PBHs will emit Hawking radiation with a temperature
$T_{h}$.  This leads to a finite PBH lifetime, as previously discussed.  Further, 
the Hawking Radiation pressure:
\begin{equation}
p_{h} = \sigma_{SB}^* T_{h}^4
\end{equation}
where $\sigma_{SB}^*$ is the (dimensionless) modified Stefan-Boltzmann constant for Hawking radiation.  This leads to a radiation force of
\begin{equation}
F = p_{h} \left(\frac{R_S}{2D}\right)^2 \pi r_{acc}^2 = \left(\frac{9\sigma_{SB}^*}{2^{20}\pi^3}\right)M_p^2 \left(\frac{M_P}{M}\right)^2 \left(\frac{R_S}{D}\right)^2
\end{equation}
And a timescale for orbital disruption of
\begin{equation}
t_{hawk} = \left(\frac{2^{31/2}\pi^3}{9\sigma_{SB}^*}\right) M_p^2 R_S^3 \left(\frac{D}{R_S}\right)^{3/2};
\end{equation}
note this is different from the PBH lifetime $\tau$;
\begin{equation}
\tau = \left(\frac{2^5 \pi^3}{3 \sigma_{SB}^*}\right) M_p^2 R_S^3
\end{equation}
So that
\begin{equation}
\frac{t_{hawk}}{\tau} = \frac{2^{21/2}}{3} \left(\frac{D}{R_S}\right)^{3/2} \gg 1.
\end{equation}
The conclusion being that the PBH will completely evaporate well before the evaporated radiation 
pressure can disrupt the binary.

\subsection{Gravitational Radiation} Finally, For close binaries of PBHs, the loss of energy due to 
gravitational radiation may become important.  We estimate the size of this effect as 
follows.  Consider two identical mass PBHs separated from their center of mass by a distance $D$ in a circular
orbit.  The emitted power due to gravitational waves is
\begin{equation}
P = \frac{2}{5} \frac{G^4 M^5}{R^5}.
\end{equation}
Given a gravitational binding energy $U = GM^2/(2D)$, the time-scale for inspiralling due to
gravitational wave losses is 
\begin{equation}
t_{spiral} = \frac{U}{P} = \frac{5}{4} D \left(\frac{D}{GM}\right)^3
\end{equation}
This can roughly be modelled as an additional drag force;
\begin{equation}
F_{wave} = \frac{P}{v} = \frac{M_P^2}{5\cdot2^{5/2}} \left(\frac{R_S}{D}\right)^{9/2}
\end{equation}

Gravitational wave emission will cause the binary orbit to decay, and accelerate the PBH merging process.
Comparing to the orbital timescale,
\begin{equation}
\frac{t_{wave}}{t_{orbit}} = \frac{5\sqrt{2}}{4\pi} \left(\frac{D}{R_S}\right)^{5/2}
\end{equation}
For $D > R_S$, $t_{wave} \gg t_{orbit}$ and, as expected, many orbits occur before inspiraling.

\subsection{Conditions for Binary Merger}

Therefore, the only mechanism that will affect the PBH binaries will be gravitational wave emission.
The concern now is whether the PBHs can evaporate before the inspiral is complete.  Comparing the
relevant timescales,
\begin{equation}\label{inspiralcondition}
\frac{t_{spiral}}{\tau} = \frac{15 \sigma_{SB}^*}{2^4 \pi^3} \frac{G D^4}{R_S^6} \approx (0.166)\left(\frac{M_P}{M}\right)^2 \left(\frac{D}{R_S}\right)^4
\end{equation}
If $t_{spiral}/\tau < 1$, then the PBH binary will merge before evaporation takes place.  This limit
can be turned into a maximum value of the initial separation $D$ for which merging will occur.  To gauge 
what importance this effect will have, we need to know more about the distribution of initial separations.

First, define two quantities.  From the definition of $\delta(\vec{x})$ (equation \ref{delta}), $\rho(\vec{x}) = \bar{\rho} (1+\delta(\vec{x}))$.  Consider the quantity
\begin{equation}
\langle \rho(\vec{x}) \rho(\vec{x} + \vec{r}) \rangle = \bar{\rho}^2 (1 + \xi(\vec{r})),
\end{equation}
which can be shown by direct substitution and noting that $\langle \delta(\vec{x}) \rangle = \langle \delta(\vec{x} + \vec{r}) \rangle = 0$.  Assuming isotropy and a monochromatic mass function, $\rho_{PBH} = n_{PBH} M$, and
\begin{equation}
\langle n_{PBH}(\vec{x}) n_{PBH}(\vec{x}+\vec{r}) \rangle = \bar{n}^2 (1 + \xi(r)).
\end{equation}

The mean initial separation we define as $\bar{D} = \bar{n}_{PBH}^{-1/3}$.  This, being the first moment
of the density distribution, tells us nothing about clustering.  Define instead $\tilde{D} = \langle n_{PBH}^2(0) \rangle^{-1/6}$, which has the dimensions of length.  
Using equation \ref{politzer} in the limit
that $\nu >>1$ and $r \approx 0$ (as we are interested in the clustering at short distances), we find
\begin{equation}
\langle n_{PBH}^2(0) \rangle = \bar{n}_{PBH}^2 e^{\nu^2}
\end{equation}

Using Equation (\ref{beta}) and the definition of $n_{PBH}$,
\begin{equation}
\bar{D} =  \sqrt{2\pi} [N_*(\nu)]^{-1/3} e^{\nu^2/6} R_H
\end{equation}
so that
\begin{equation}
\tilde{D} = \langle n_{PBH}^2(0) \rangle^{-1/6} = \bar{D} e^{-\nu^2/6} = \sqrt{2\pi} [N_*(\nu)]^{-1/3} R_H
\end{equation}

While the number density is exponentially suppressed with $\nu$, there is a cancellation computing
this clustering length so that it is roughly constant with $\nu$.  As expected from the peak-background
split model, the scale that sets the clustering is the horizon size at PBH formation.  

\begin{figure}
\includegraphics[width=5.75in]{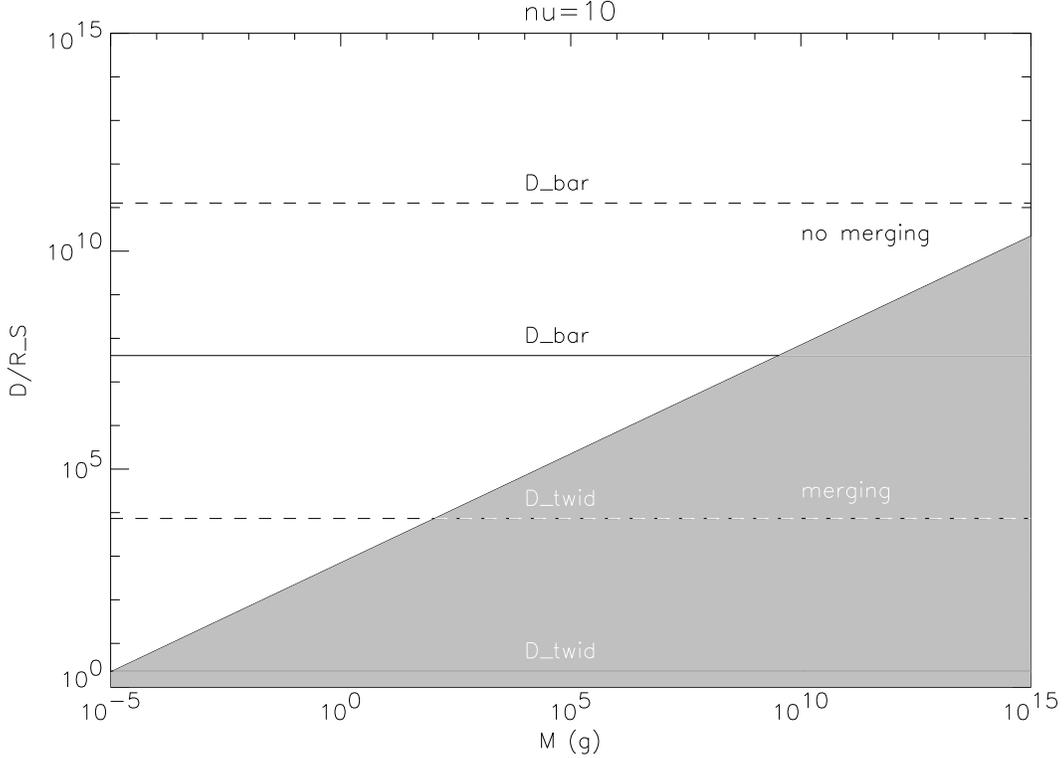}
\caption[Conditions for Binary Merger before Evaporation]{Conditions for binary merger before evaporation.  $D_{bar}$ is the mean separation assuming a uniform (non-clustered) PBH distribution.  $D_{twid} = \left(\frac{\pi}{2}\right)^{2/3}\nu^{1/3} f^{-1} R_S$ is the ``clustering length'' scale.  Solid curves are for $f=1$, dashed curves are for $f=10^{-3.5}$.\label{fig_spiral}}
\end{figure}

Taking $D=\tilde{D}$ in Equation \ref{inspiralcondition} for $f=1$ and $N_*(\nu) = \sqrt{\frac{2}{\pi}} \nu^{-1}$, we obtain
\begin{equation}
 \frac{t_{spiral}}{\tau}=2.66 \nu^{4/3} \left(\frac{M_P}{M_H}\right)^2,
\end{equation}
meaning, provided $M_H > M_P$ and with only slight $\nu$ dependence, inspiral typically occurs well before Hawking evaporation completes.
In Figure \ref{fig_spiral} we plot the conditions for merging before evaporation for the range of 
hole masses expected to evaporate by the current day.  We see that, assuming binary formation occurs,
the PBH clustering implies certain merging over most of this range, even relaxing the assumption on $f$.

Let us talk a bit more about this result.  For PBHs with $M < 10^{15}$g, this implies that if the formation
of PBH binaries is efficient, the lifetime of the PBH population as a whole is increased, as $\tau \propto M^3$.
Given hierchical clustering, where a new generation of PBH binaries are formed from the first generation, 
this process can theoretically proceed until all smaller PBHs are bound up into larger ones.  

This would have
a profound impact upon cosmology -- as PBHs that should have evaporated by now would still be present in the
universe.  There are two caveats in this scenario.  First, if PBH binary formation is inefficient, then most PBHs
will still evaporate ``when they should'' and the merged ones have little impact.  Second, it assumes that
the PBH binaries are not disrupted by other means; such as other PBHs.  We address this point next.

\subsection{PBH cluster occupation}\label{PBHCO}
Having shown that gravitational wave emission can cause PBHs to merge before they evaporate, the
creation and destruction of PBH binaries in a PBH cluster will determine whether merging is a common
enough even to impact the PBH mass function as a whole.  

We define a PBH cluster as a group of PBHs that are within causal contact (sub-horizon)
and are gravitationally bound.  
%%Just as with the formation of LSS, PBH clusters are
%%expected to undergo hierarchical clustering and mergers as time progresses.  Because of this,
%%this topic is difficult to examine analytically, while numerical simulations of N-body systems
%%have had much success in reproducing observed galaxy distributions.
We showed in \citep{chisholm} that, under the assumption that $\nu \gtrsim 4$, PBHs are distributed at large distances as 
clusters of objects with mean occupation number
\begin{eqnarray}\label{N_c}
N_c & \sim & N_*(\nu) e^{\nu^2/4}.
\end{eqnarray}

The question then remains: what range of parameters ($M_0$, $\nu$) are physically interesting, given the absence of 
evidence for PBHS in our current universe?  In \citep{chisholm}, we constrained these according to the criteria that 
a) PBHs do not induce an early matter-dominated phase of the universe, and b) PBHs, as dark matter, do not produce a 
too large isocurvature perturbation.  These give a parameter space where the allowed $\nu$ decreases as $M_0$ increases
(see their Figures 1-3).

We can then look at two different categories of PBHs, distinguished by mass: those that would and would not have 
evaporated by the current day, with the boundary at $M_0 \sim 10^{15}$ g.  The subsequent evolution of this first 
category (the lightest PBHs) is interesting in that the evaporation lifetime $\tau \propto M^3$ -- if merging occurs,
this will increase the PBH lifetime, perhaps above the ``evaporation'' boundary.  For $M_H < 10^{15}$ g, this gives a range 
of $\nu \gtrsim 7-10$ for $f = 10^{-3.5}$ and $\nu \gtrsim 8-11$ for $f=1$.  For $M_H > 10^{15}$ g there is no worry 
of Hawking evaporation, but these have a smaller lower bound on $\nu$ and thus would form (on average) smaller clusters.  

The evolution of bound PBH clusters will depend upon their internal density profile.
The average cluster profile is just $\langle\rho_{PBH}(r)\rangle = \rho_{PBH} \left(1+\langle\delta_{PBH}(r)\rangle\right)$.  
Since a randomly chosen point is likely to be located at a peak, it can be shown that\citep{peacock}:
\begin{equation}
 \langle\delta_{PBH}(r)\rangle = \frac{\delta_{PBH}(0)}{\xi_{PBH}(0)} \xi_{PBH}(r) \propto \xi_{PBH}(r).
\end{equation}
The cluster is highly peaked, due to the enhancement of the PBH correlation function.  The exact details of the shape
of the cluster will depend upon choice of initial radiation power spectrum $P_{rad}(k)$, window function $W_k(kr)$ and
higher order statistics (we are only considering the 2-point function).  Nonetheless, while PBHs can form no closer 
than a horizon distance apart, their typical separation within the cluster $\tilde{D}$ cannot be too much larger than
this distance due to the exponential enhancement of the correlation function for small $r$.  Due to the high central
density (below we compute the concentration parameter $c$), this results in the PBHs being approximately "close-packed" 
inside the core.  Outside the core, the density falls off faster than the underlying radiation perturbation, as 
$\xi_{PBH}(r) = \exp\left(\nu^2 w(r)\right) - 1$.  Far from the core, the cluster profile falls off no slower than the 
underlying radiation perturbation profile: as $r \rightarrow \infty$, $\xi_{PBH}(r) \approx \nu^2 w(r)$.  The steepest
profile comes from assuming $P_{rad} \propto \delta_{\textrm{Dirac}} (k-k_*)$ (a delta function a given scale $k_*$, which
we can take to be the horizon scale at formation); 
there $w(r) = \frac{\sin(k_* r)}{k_* r}$ and asymptotically $w(r) \propto r^{-1}$ as $r \rightarrow \infty$.  In this
extreme limit, normalizing the cluster profile $\langle\rho_{PBH}(r)\rangle$ to the mean objects in a cluster $N_c$ puts
virtually all PBHs within $k_*^{-1}$ -- this would be unphysical give this is roughly the event horizon size, but motivates
our later assumption of putting all of the PBHs within the core.   A more
realistic power spectrum flattens this profile; taking $P_{rad} \propto k^n$ gives $w(r) \propto r^{-(n+3)}$ 
asymptotically\citep{peacock}.  

The quasi-equilibrium state of an N-body system of gravitating point masses can be broken down into two major components:
a high density central core and a low-density encompassing halo \cite{saslaw}.  Due to the enhanced clustering, the 
PBH clusters considered here already begin with a similar, centrally concentrated profile.  
It is convenient to discuss separately activity within the core (small radius, or $r \rightarrow 0$) and in the halo (large radius, or $r \rightarrow \infty$) of the cluster.  
The core concentration $c$ is then
\begin{equation}
 c = \frac{\rho_{core}}{\rho_{halo}} = \frac{1+\frac{\delta_{PBH}(0)}{\xi_{PBH}(0)}\xi_{PBH}(0)}{1+\frac{\delta_{PBH}(0)}{\xi_{PBH}(0)}\xi_{PBH}(\infty)} \approx \frac{\xi_{PBH}(0)}{1} = \sigma_{PBH} \approx e^{\nu^2/2}.
\end{equation}
This alone is a clue that our cluster is destined for collapse; an isothermal sphere with $c > 708.61$ is unstable to collapse 
(known as the gravothermal catastrophe) \cite{LBW}.  This happens for $\nu > 3.62$, which is in the parameter space we were considering anyway.
Further, as the core begins to contract and PBHs begin to approach relativistic speeds,
the relativistic instability \cite{zeldovich3} accelerates this collapse.  To confirm this, we examine in more detail the processes involved
in this collapse process.

As our clusters begin ``pre-collapsed'' in a sense, there is expected to be a great difference between core radius and halo radius when it comes
to estimating the extent of the clusters.  Due to its high density, it is within the core that most interactions take place \cite{spitzerbook, saslawbook}, and
so it is within the core we are more interested anyway.  We can make some estimates of size using our expressions 
for $N_c$ and $\tilde{D}$, assuming the majority of PBHs form within the core.  As $N_c \propto V_{core} \propto R_{core}^3$,
the cluster radius $R_{core} \propto N_c^{1/3}$.  In the core, where the PBHs are most concentrated, each PBH is separated by a typical distance of
$\tilde{D}$, then
\begin{equation}
R_{core} \sim N_c^{1/3} \tilde{D} = \left(N_*(\nu)e^{\nu^2/4}\right)^{1/3} \times \sqrt{2\pi} \left(N_*(\nu)\right)^{-1/3} R_H = \sqrt{2\pi} e^{\nu^2/12} R_H.
\end{equation}

Thus, the core size is independent of our choice of $N_*(\nu)$, but increases with increasing $\nu$, as expected. 
Taking a range of $2 < \nu < 10$, this gives a range of $3.5 R_H < R_{core} < 10^4 R_H$.  As the higher values of
$\nu$ are only allowed for PBHs that form earlier, these large cores will consist of smaller mass PBHs (and a smaller 
horizon size), while smaller cores will have larger mass PBHs.

We can similarly make an estimate for the initial size of the surrounding halo.  Recall that the mean separation of PBHs is $\bar{D} = \bar{n}_{PBH}^{-1/3}$.
As we are now grouping PBHs into clusters of size $N_c$, the density of clusters of PBHs is smaller an the density of PBHs themselves, thus the 
mean separation between clusters $D_c$ must be greater than the mean separation between PBHs themselves: 
\begin{equation}
D_c = \left(n_{PBH}/N_c\right)^{-1/3} = N_c^{1/3} \bar{D} \sim \left(N_*(\nu)e^{\nu^2/4}\right)^{1/3} \times \sqrt{2\pi} \left(N_*(\nu)\right)^{-1/3} e^{\nu^2/6} R_H = \sqrt{2\pi} e^{\nu^2/4} R_H
\end{equation}
This sets an upper limit on the halo radius, imagining the PBH clusters close-packing in space.  As the evolution of the cluster is driven primarily 
by core activity, we will focus on that from here on out.

%%We now turn to the behavior of systems with large $N$.  For this,
%%we assume that all interactions that take place to be subsets of $N=2$ (collisions, scattering and binary formation) 
%%and $N=3$ (flyby, ionization and resonance) body interactions (see \cite{binney} for more details on these).  The 
%%interaction rate of an $N$ body process scales as $\rho^{N}$; so most of the behavior of a system will be
%%governed by $N=2$ interactions, with higher order interactions only comparable in high density areas.

Take a PBH cluster of $N$ initial PBHs of mass $M$ with scale radius $R_c = s R_S$, where $R_S = 2 GM$ 
is the initial Schwarzschild radius.  The dynamical timescale (roughly the cluster crossing time) is
\begin{equation}
t_{c} \sim \sqrt{\frac{R_c^3}{GM_c}} = \sqrt{\frac{2}{N}} s^{3/2} R_S.
\end{equation}
 
As a reference, the binary period and inspiral timescales are
\begin{equation}
t_{orbit} = 4\pi \sqrt{2} d^{3/2} R_S,
\end{equation}
\begin{equation}
t_{spiral} = 10 D \left(\frac{D}{R_S}\right)^3 = 10 d^4 R_S,
\end{equation}   
where $d = D/R_S$.

The first process we need to consider for large N-body systems is that of relaxation.  This is the
process by which a cluster achieves equilibrium (virialization) through the combined effect of 
two-body scatterings.  The relaxation timescale is given by \citep{spitzer, binney}
\begin{equation}
t_{rel} \sim \left(\frac{0.14 N}{\ln (0.4 N)}\right) t_{c}.
\end{equation}
Gravitational wave emission during 2-body scatterings will accelerate cluster relaxation (so that the expression for $t_{rel}$
is an upper limit), in addition allowing for binary mergers.  Under the assumption of violent relaxation, however, the cluster virializes in only a few
dynamical times.  After virialization, the PBHs will have a velocity dispersion
\begin{equation}
\langle v^2 \rangle \sim \frac{GM_c}{R_c} = \frac{N}{2s}
\end{equation}

The simplest mechanism for PBH mergings is direct collisions; the timescale being
\begin{equation}
t_{coll} \sim 0.8 \ln (0.4N) \left(\frac{\Theta^2}{1 + \Theta}\right) t_{rel}
\end{equation}
where $\Theta = (4 v^2)^{-1}$.  This process is initially negligible compared
to the relaxation time for small ($N \lesssim 100$) clusters.

Successive scatterings can give a single object enough energy so that it can escape from the 
cluster entirely ($v > v_{escape}$), while the cluster shrinks in size.  This mass loss gradually
leaves to the ``evaporation'' of the cluster; where most of the bodies are ejected to infinity, 
leaving behind only a hard binary system (in the absence of gravitational wave emission) or a central 
black hole (where gravitational wave emission has induced orbital decay and merging of the hard binary).
The timescale for this evaporation
is
\begin{equation}
t_{evap} \sim 300 t_{rel}.
\end{equation}

The cluster shrinking is accompanied by core collapse where the innermost portions of the cluster accrete more
and more of the mass of the cluster, which could result in runaway growth unless halted by some mechanisms.  Numerical
simulations \cite{kupi, lee} show that this process begins within 10-20 relaxation times -- much smaller than the cluster
evaporation timescale.

\subsection{Binary Formation}

We now turn to binary formation in a cluster.  This is a critical point, because a population of 
binaries in the core could eject other PBHs from the core while themselves contracting in their orbit (binary hardening).
This, effecively, cools the core, and could possibly halt core collapse (see \citep{binney} for a discussion).  This is the
case when PBHs are moving non-relativistically (in the Newtonian regime).  Once relativistic effects are considered, in
particular gravitational wave emission, we see that this is not the case, and binaries are not enough to arrest core collapse.

Since the PBHs form at rest with respect to the cosmic expansion, there are no binaries present at the ``birth'' of the 
cluster \footnote{Known as primordial binaries, where primordial is being used in a similar context.}.  In the 
Newtonian regime (cluster members moving non-relativistically), the only avenue for binary formation is through
3-body exchange interactions only, with a formation timescale of
\begin{equation}
t_{3} \sim 10 N^2 \ln N t_{rel}
\end{equation}
As cluster ``evaporation'' occurs in a few hundred relaxation times, binary formation through
3-body interactions is negligible for large $N$ systems ($N \gtrsim 100$).  

Binaries are only formed through 2-body interactions when there is some energy dissipation involved during (what would have been) a scattering event.
For stars, this is through dissipation in the stellar atmospheres.  For black holes, this happens due to gravitational wave emission during 
the scattering, which heretofore we have not considered.  This has been studied in detail for compact clusters of black holes by a number of authors:  
in \cite{QS1, QS2}, analytic esimates and Fokker-Planck simulations are presented, while N-body simulations are presented in \cite{lee, kupi}.
While their simulations were for stellar-mass black holes, the results are mass independent.  There, 2-body formation is dominant over 3-body formation, with
the ratio of timescales (from \cite{QS1}):
\begin{equation}
 \frac{t_2}{t_3} = \frac{300}{N} v^{15/14}.
\end{equation}
At the outset of cluster contraction, $v \ll 1$ and 2-body binary formation is dominant for $N > 300$ clusters.  Only as
the core evolves can $v \rightarrow 1$ (i.e., becomes relativistic) and $N$ shrink enough for 3-body formation to become important.  Despite the enhanced
formation in the core, the inspiral time $t_{spiral} < t_{rel}$ is smaller than the cluster relaxation time, so that these
binaries are not effective at heating the core and halting collapse (noted in \cite{QS1}).

%See Figure \ref{fig_timescales}
%to see these various timescales as a function of $N$.

%\begin{figure}
%\includegraphics[width=5.75in]{./timescales.eps}
%\caption[Timescales for Cluster Evolution]{Timescales for cluster evolution.  Curves are labeled as in the text; assumes $\nu=10$, $s \approx 10^5$, $d=\tilde{D}/R_S$, $f=0.1$.  \label{fig_timescales}}
%\end{figure}

%From energy arguments, as bodies leave the system through evaporation and ejection, the cluster
%left behind must shrink (increased potential energy causes contraction).  This contraction leads to
%a process known as core collapse within a score of relaxation times, where the central density, 
%analytically, becomes infinite \citep{binney}.  For the case of PBHs, this is the physical case
%as well, as PBH merging accelerates as the central density increases (due to head-on collisions as well as
%binary inspiralling).  Since core collapse occurs well before the cluster could gravitationally evaporate, one might
%expect only a minority of the initial PBHS are able to escape the cluster without being captured into the central, core
%BH that would develop.  What remains of the initial PBH cluster is a single, more massive, BH.

Just what mass fraction $f_{core}$ of the initial cluster ends up in a central black hole is not well known.  Numerical calculations by \cite{kupi, lee} 
find $f_{core} \sim 0.06 - 0.1$, though the simulation begins to break down at that point, meaning this is a lower limit.  Of those PBHs not captured in the core (either
ejected from the cluster or remaining in the halo), they will either Hawking evaporate (if they are light enough) or remain until the
universe becomes matter-dominated and they are bound up within galaxies.  The possibility remains that some residual number of PBHs from 
this process might survive until the current day, providing an observational test of this scenario.

%%%%%%%%%%%%%%%%%%%%%%%%%%%%%%%%%%%%%%%%%%%%%%%%%%%%%%%%%%%%%%%%%%
%%%%%%%%%%%%%%%%%%%%%%%%%%%%%%%%%%%%%%%%%%%%%%%%%%%%%%%%%%%%%%%%%%
\section{Conclusions}\label{conc}
%%%%%%%%%%%%%%%%%%%%%%%%%%%%%%%%%%%%%%%%%%%%%%%%%%%%%%%%%%%%%%%%%%
%%%%%%%%%%%%%%%%%%%%%%%%%%%%%%%%%%%%%%%%%%%%%%%%%%%%%%%%%%%%%%%%%%

A consequence of the PBH clustering, developed in the previous section, is the merging of PBHs into more
massive, longer lived PBHs during the radiation dominated era.  This implies that PBHs are strong candidates 
to be the ``seed'' BHs that form
SMBHs.  This is distinct from other models in the literature \citep{bean, duchting, custodio1} where
PBHs serve as SMBH ``seeds'', which are of two types.  First \citep{bean, custodio1}, PBHs formed from a 
blue spectrum of perturbations undergo hierarchical merging along with DM halos in the MD regime.  It was found,
however, that it is still difficult to account for the observed BH mass without invoking some additional
accretion source; in this case the accretion of a cosmological quintessence field.  Quintessence fields are
types of scalar fields that 
arise in theories of dark energy, and unlike a cosmological constant, BHs are able to accrete energy from a 
time-dependent scalar field \citep{jacobson}.  The second type of PBH seed theory \citep{duchting} assumes that
the $\sim 1000 M_\odot$ seed BHs are PBHs formed at just that right mass scale.  This requires a deviation from
scale-invariance in the power spectrum very close to the era of BBN, which is highly constrained.

The advantage of our model is that PBH formation occurs much earlier than in \cite{duchting}, so that the
power spectrum isn't as constrained.  Further, PBH merging takes place in the RD epoch, so there is more time
for PBHs to merge, and possibly no need to invoke an additional accretion mechanism as in \cite{bean, custodio1}.
A more detailed study of this model is planned to determine whether this is the case.

PBH merging in clusters dramatically changes the limits on initial PBH abundance $\beta$.  
The rate of merging is sensitive to the initial conditions of the cluster.  We have considered
an idealized scenario, with PBH formation happening at a single mass scale and at a single time.
Physically, one would need to account for PBH creation across a span of times, and include in
the cluster dynamics the effects of a spectrum of masses.  N-body simulations would need to be 
carried out to examine this further.

In addition to providing the seeds of SMBHs, this PBH merging scenario we have discussed has
other predictions.  One prediction is more gravitational wave emission than originally assumed
for a uniform PBH population.  This is due to the increased probability of PBH binary formation
and emission from $N>2$ bound states, specifically in the core of a PBH cluster.  This would alter
recent predictions of gravitational wave spectra from direct graviton emission from PBHs \cite{saito, anantua}.

\bibliography{part2}

\end{document}